\documentstyle[12pt,epsf]{article}

\begin{document}
\newcommand {\be}{\begin{equation}}
\newcommand {\ee}{\end{equation}}
\newcommand {\bea}{\begin{array}}
\newcommand {\cl}{\centerline}
\newcommand {\eea}{\end{array}}
\renewcommand {\theequation}{\thesection.\arabic{equation}}
\newcommand {\newsection}{\setcounter{equation}{0}\section}

\baselineskip 0.65 cm
\begin{flushright}
IPM/P-99/20 \\
hep-th/9903107
\end{flushright}
\begin{center}
 {\Large {\bf One Loop Renormalizability of Supersymmetric Yang-Mills 
Theories on Noncommutative Two-Torus}}
\vskip .5cm

 M.M. Sheikh-Jabbari
\footnote{ E-mail: jabbari@theory.ipm.ac.ir } \\

\vskip .5cm

 {\it Institute for studies in theoretical Physics and mathematics 
IPM,

 P.O.Box 19395-5531, Tehran, Iran}
\end{center}

\vskip 2cm
\begin{abstract}
We argue that Yang-Mills theory on noncommutative torus, expressed in the
Fourrier modes, is described by a gauge theory in a usual commutative space,
the gauge group being a generalization of the area preserving diffeomorphisms
to the noncommutative case. In this way, performing the loop calculations in
this gauge theory in the continuum limit, we show that this theory is {\it 
one loop renormalizable}, and discuss the UV and IR limits. The moduli 
space of the vacua of the noncommutative super Yang-Mills theories in 2+1 
dimensions is discussed.

\end{abstract}
\newpage
\section{Introduction}

It has been pointed out by A. Connes, M. Douglas and A. Schwarz (CDS) [1] 
that the supersymmetric Yang-Mills theory (SYM) on noncommutative torus (NCT), 
is naturally related to compactification of Matrix theory with a constant 
$C_3$ field. 
Since the eleven dimensional three-form flux is related to the NSNS
two-form flux in string theory, CDS's conjecture means that the dynamics of 
D-branes in a B-field background is described by gauge theory on a 
noncommutative torus, where the B-field gives the deformation parameter 
of the torus, in the zero volume limit [2,3,4,5,6,7,8].

The noncommutative torus (NCT) is a basically flat and compact space in which 
the coordinates do not commute:
\be
[x^i, x^j]={\theta^{ij} \over 2\pi i}.
\ee
In this paper we mostly consider the two-torus and hence in our case the
NCT is described by only one parameter $\theta$ which is called deformation 
parameter of the torus [9].

It was first explicitly shown in [4], completed and generalized in [7,10],
that the above commutation relation naturally appears for the components
of open strings attached to D-brane in the B-field background. Further 
study on these open string dynamics and quantizing them revealed that
their low energy dynamics is governed by the noncommutative supersymmetric 
Yang-Mills (NCSYM) theory defined on the NCT [2,11].

Since NCSYM describes the low energy interactions of open strings 
at which it is decoupled from string theory dynamics and some
special limits of M-theory, it is believed that NCSYM should be a well-defined
quantum theory [1,11,12].
But, as pointed out in many papers e.g. [1,2,13,14], NCSYM is a
{\it non-local} field theory, where the non-locality scale is identified with
the deformation parameter of the torus which is a {\it dimensionless} parameter
and hence this non-locality is expected to remain at all energies; and the 
question of renormalizability, renormalization flow and existence of 
fixed points of the non-local field theories has not been answered yet. 
Existence of a UV fixed point, in which the theory is scale invariant, is one 
of the elements that make a usual field theory, e.g. a gauge theory, work. 
Using this fixed point, we are able to define the theory in the continuum limit,
but in the NCSYM, since it is a non-local field theory, the concept of 
renormalization and fixed points should be understood and studied in some other
sense.

In this paper, first we briefly review the algebraic structure of NCT and
the gauge bundles defined on it, and discuss the Morita 
equivalence between sections of these bundles on different tori. 
This equivalence enables us to map a $U(N)$ NCSYM with 
magnetic flux $M$ to a $U(1)$ NCSYM. So, having studied the $U(1)$ NCSYM,  
we can discuss the $U(N)$ case on the same footing.

In section 3, by using the Fourrier transformation on the NCT, we map the  
$U(1)$ NCSYM to a conventional commutative gauge theory, with a gauge 
group which is a generalization of the area-preserving diffeomorphisms
of the two-torus, SDiff($T^2$), to the noncommutative case.
We extensively study the propagators and vertices of this gauge theory.

In section 4, performing the explicit loop calculations we show that, 
the related divergences up to one loop are like a usual gauge theory, 
in which the quadratic Casimir of the group, $C_2(G)$, is ${2}$.   

In section 5, we argue that NCYM in UV and IR limits behaves as in a usual 
gauge theories. In other words, although NCYM is a non-local theory, 
the theory is one loop renormalizable and it admits a UV fixed point.   
We also briefly discuss the moduli space of vacua and R-symmetries of 
NCSYM, as a theory with 16 real super-charges. 

Finally in section 6, we will summarize our results and discuss some open 
and unanswered issues on this problem.

\section{Noncommutative Yang-Mills Theory, A Review}
\setcounter{equation}{0}

In order to build the YM theory on a noncommutative torus, we briefly review
the necessary ingredients from noncommutative geometry.
For an extensive reference, we refer the reader to Connes' book [9]. 
\newline
{\it Gauge Bundles on Noncommutative Torus}
\newline
According to noncommutative geometry formulation, geometric spaces are 
described by a $C^*$-algebra which is not in general commutative. This is the
generalization of the Gelfand-Nimark theorem which substitutes a compact 
manifold by the algebra of functions on it. In the case of noncommutative 
two-torus, the related $C^*$-algebra can be generated by the Fourrier
modes:
$U_i=e^{2\pi i x^i}, i=1,2$, where
\be
U_1U_2=e^{2\pi i \theta}U_2U_1.
\ee
or equivalently $\;\;\ [x^1,x^2]={\theta \over 2\pi i}$.
The same algebra can also be generated by the star product among these  
functions:
\be
(f*g)(x)=exp{(\pi i \theta {\epsilon^{ij}}{\partial \over \partial x^i}
{\partial \over \partial y^j})} f(x) g(y)|_{y=x}.
\ee
In the above definitions $\theta$ is the deformation parameter of the NCT. 
We also need a set of derivatives $\partial_i$, which satisfy
\be
[\partial_i,\partial_j]=0\;\;\;\;, \;\;\;\; [\partial_i,x^j]=\delta_i^j.
\ee
By generalizing the Serr-Swan theorem to the noncommutative spaces, gauge 
bundles are the finitly projective right (left) -modules over the $C^*$-algebra. 
So one can realize the connection, $\nabla_i$, over a bundle by: 
\be 
\nabla_i (c A)= (\nabla_i c) A + c\partial_i A, 
\ee
where $c\in C^* \; , \; A\in$projective right-module over $C^*$. According 
to (2.4) the gauge connection commutes with all of the elements in $C^*$
and hence it is not a function of noncommutative $x^i$, but there is  a
combination of $x^i,\partial_j$ [8], 
\be
\tilde{x}^i=x^i+{i \theta \over 2\pi}\epsilon^{ij}\partial_j,
\ee
which do commute with $x^i$ and the gauge bundles and the related gauge 
transformations should be a function of $\tilde{x}^i$: 
\be
\nabla_i=\partial_i +i A_i(\tilde{x}),
\ee
and the gauge fields $\tilde{x}$, can be expanded by Fourrier modes
$\tilde{U_i}=e^{2\pi i \tilde{x}^i}$;
\be
\tilde{U_1}\tilde{U_2}=e^{-2\pi i \theta}\tilde{U_2}\tilde{U_1}.
\ee
Although up to here, we have discussed only the $U(1)$ bundles, the $U(N)$ 
generalization is trivially possible, by simply letting $A$ take values in
the corresponding Lie-algebra.

It is interesting to construct the parallel string theoretic description of 
the mathematics we have  built here. According to Witten [15], 
the zero modes of the open strings attached to a D-brane form
a gauge multiplet of the $U(1)$ theory living on the brane, and these gauge
fields are functions of the zero modes the corresponding 
open strings. In the presence of a non-zero B-field background, in spite of 
the noncommutative coordinates, this is still valid. To see this let us  
consider the mode expansion of such open strings [7,10,11]
\be
X^{i}=x^i+p^i \tau-B^i_j p^j \sigma + Oscil. \;\ i,j=1,2,
\ee
where $p^i$ are the related momentums and $x^i$ are an arbitrary point on the 
tours. Quantizing these open strings, we find:    
\be
[p_i,p_j]=0\;\; , \;\; [p_i,x^j]=-i\delta_i^j \;\; , \;\; [x^i,x^j]=iB^{ij}.
\ee
In the above algebra $p_i$ can be represented by $-i\partial_i$. The  
coordinates of these open strings at $\sigma=\pi$ which form the corresponding 
gauge multiplet, is $\tilde{x}^i$ (2.5). 

Following  't Hooft, we can construct the non-Abelian $U(N)$ gauge bundles  
with a non-zero magnetic flux $M$ \footnote{We assume $M,N$ that are 
relatively prime.}. Consider $U(N)$ matrices:
\be 
V_1V_2=e^{2\pi i {M\over N}}V_2V_1.
\ee
The generalization of $\tilde{U_i}$, the Fourrier expansion basis, to this case
can be obtained by $Z_i$:
\be
Z_1=e^{{2\pi i \over N} \tilde{x^1}} V_2^K \; , \; 
Z_2=e^{{2\pi i \over N+M\theta} \tilde{x^2}} V_1^{-K},
\ee
where  $K$ is an integer satisfying  $NL-MK=1, L\in Z$. one can easily check that
\be
{Z_1}{Z_2}=e^{-2\pi i \hat{\theta}}{Z_2}{Z_1}\;, \; \hat{\theta}={K+L\theta \over
N+M\theta}.
\ee
Along the lines of [8], we can build the gauge connection through $Z_i$: 
\be
exp \nabla^0_1=e^{\partial_1} V_1 Z_2^{-M}\; , \; 
exp \nabla^0_2=e^{\partial_2} e^{-2\pi i {M\over N}\tilde{x}^1}V_2 Z_1^M, 
\ee
and
\be
\nabla_i=\nabla^0_i+ i A_i(Z).
\ee
The most important feature of the above gauge bundle construction is the 
Morita equivalence, which is a mathematical equivalence between the
bundles on the different noncommutative tori. Comparing (2.11) and (2.1),
we see that both $\tilde {U_i}$ and $Z_i$ generate the same type of algebra;
the corresponding deformation parameters are related by an Sl(2,Z) 
transformation, (2.12). Hence, we can identify the {\it non-Abelian}  
gauge field $A$, with an {\it Abelian}  one, $\hat{A}$, which lives on another
NCT defined by $\hat{\theta}$. This equivalence between the gauge bundles
on different noncommutative tori, is a special form of {\it Morita equivalence}.

In writing the  explicit form of the NCSYM, we prefer to use the algebra of 
functions defined by star product, (2.2). According to CDS [1],
the noncommutative $U(1)$ gauge connection can be built by
\be
\nabla_i=\partial_i+i\{A_i,\;\;\ \}_{M.B.}
\ee
with 
\be\bea{cc}
\{f,g\}_{M.B.}(x)=(f*g)(x)-(g*f)(x)\\
\;\;\;\;\;\;\;\;\;\;\;\;\;\;\;\;\;\;\;\;\;\;\;\  =2i f(x) \sin{(i\pi \theta 
\epsilon^{ij}\stackrel{\leftarrow}{\partial_i}\;
\stackrel{\rightarrow}{\partial_j})} g(y)|_{y=x}.
\eea\ee
The curvature is
\be
F_{\mu\nu}=[\nabla_{\mu},\nabla_{\nu}]=\partial_{[\mu}A_{\nu]} +
i\{A_{\mu},A_{\nu}\}_{M.B.} \;\;\; \mu=0,1,2. 
\ee
Then $U(1)$ NCYM (on $R\times T^2_{\theta}$) is given by
\be
S={1\over g^2_{YM}}\int d^3x \; F_{\mu\nu}F^{\mu\nu}.
\ee
The above action enjoys the gauge invariance:
\be
A_{\mu}\rightarrow A_{\mu} +\partial_{\mu}\epsilon+i\{\epsilon,A_{\mu}\}_{M.B.} 
\ee
We can also supersymmetrize this action by adding the correct fermionic and 
scalar degrees of freedom [5,16]. The maximally supersymmetric action
which has 16 
real super-charges is of the form:
\be\bea{cc}
S={1\over g^2_{YM}}\int d^3x \; F_{\mu\nu}F^{\mu\nu}
-2 g^2_{YM} (\nabla_{\mu}X^a)(\nabla^{\mu}X^a)+ 2 g^4_{YM}(\{X^a,X^b\}_{M.B.})^2\\
-2i \Theta^{\alpha}\Gamma^{\mu}_{\alpha\beta}\nabla_{\mu}\Theta^{\beta}
+{1\over 4} g_{YM}\Theta^{\alpha}\Gamma^{a}_{\alpha\beta}.
\{X_a,\Theta^{\beta}\}_{M.B.},
\eea\ee
where $\mu,\nu=0,1,2,\; a,b=3,...,9$ and
$$
\nabla_{\mu}X^a =\partial_{\mu}X^a +\{A_{\mu},X^a \}_{M.B.},
$$
$$
\nabla_i\Theta^{\alpha}=\partial_i\Theta^{\alpha}+\{A_i,\Theta^{\alpha}\}_{M.B.}.
$$
The above action is a $D=10,\; N=1$ non-Abelian Yang-Mills theory, 
dimensionally reduced to 2+1, with the group commutators substituted for the 
{\it Moyal bracket}.

It is worth noting that, to our knowledge, there are few theories which admit  
the maximally supersymmetric extension, i.e. 16 real super-charges. 
Among these theories, those with at most two derivatives are all dimensional  
reductions of ten dimensional SYM [17]. The other example of these theories
are Born-Infeld actions. Despite lack of Lorentz invariance and higher 
derivative terms, NCSYM theories have a supersymmetric extension, 
and this property supports the renormalizability of these theories. 
Even these theories, like the SYM theories, satisfy the 
{\it non-renormalization} theorem, i.e. they are finite.

For the rational $\theta$ case, as we discussed earlier, the NCSYM, by means 
of Morita equivalence, is mapped to a conventional $U(N)$ SYM theory with a 
magnetic flux, which we know is renormalizable. For the general real
$\theta$,
we will show that by mapping the theory to the momentum space, we can analyze 
and study it in the conventional perturbative gauge theory language.

\section{NCSYM in the Momentum Space}
\setcounter{equation}{0}

Although NCSYM has been argued to be a non-local field theory [2,12,13]
with the non-locality scale remaining at all scales, we want to build another 
formulation of the theory in which the theory looks like a usual gauge theory.
Remembering (2.3) or string theoretic version of it (2.9), one observes that
expressing the NCYM in the $\partial_i$ or momentum basis, or equivalently
in the Fourrier modes, formally removes the higher derivative terms already
present in the Moyal bracket.

To see this explicitly, let us consider the Fourrier expansion of the 
gauge field $A(\tilde{x})$ [5],
\be
A_{\mu}(\tilde{x})=\sum_k A_{\mu}(\vec{k}) L_{\vec{k}},
\ee
where $\vec{k}$ is a vector on the {\it dual commutative torus}, defined in  
the usual manner, i.e. assume that $\theta$ is set to zero and dualize 
the torus, and $L_{\vec{k}}$ are the expansion basis,
\be
L_{\vec{k}}=e^{ik_i\tilde{x}^i}.
\ee
In order to analyze the NCYM theory first we study the algebra defined by 
$L_{\vec{k}}$
\be
[L_{\vec{m}},L_{\vec{n}}]=\{L_{\vec{m}},L_{\vec{n}}\}_{M.B.}=
2i \sin{({\pi\theta}\vec{m}\times \vec{n})}L_{\vec{m}+\vec{n}},
\ee
with $\vec{m}\times \vec{n}=\epsilon^{ij}m_in_j$. 

For the rational $\theta$ case, e.g. $\theta={1\over N}$, by choosing the 
proper normalization for $L_{\vec{m}}$, the above algebra is
\be
[L_{\vec{m}},L_{\vec{n}}]=
2i {N\over \pi}\sin{({\pi \over N}\vec{m}\times \vec{n})}L_{\vec{m}+\vec{n}}.
\ee
In the {\it large N}  limit (3.4) is nothing  but, the area-preserving 
diffeomorphism algebra of a usual torus, SDiff(${T}^2$), which is identified 
with $SU(\infty)$ [16]. This is a rather simple justification of the Matrix 
model M2-brane dynamics correspondence [18]. For the general rational 
$\theta$, one can check that for $\theta={q\over N}$, (3.4) again holds 
but $m_i, n_i$ are defined $mod \; qN$ and it represents a subgroup of   
$SU(\infty)$ [1,16]. In the irrational $\theta$ case, however the
situation is a little different, in this case (3.3) can not be mapped to any 
subgroup of $SU(\infty)$. 

Indeed (3.3) can be thought as the generalization  of SDiff($T^2$) to the NCT, 
SDiff($T_{\theta}^2$). In the usual geometric language, SDiff($T^2$) is the set 
of the diffeomorphisms keeping the Kahler structure of the torus unchanged, 
$\rho=i V$, $V$  the volume of the torus.
In a more general case we know that the Kahler structure can have a real part:
$\rho=i V+ b$, and as recently discussed [4,7,8,11,12,19], at least in
the zero volume limit, $b$ which in the string theory language is the 
background B-field flux, identified with the deformation parameter of the 
torus, $\theta$. One can show that under the transformations generated by (3.3) 
algebra $\rho$ is invariant. This fact supports arguments of  [7,8] about the  
U-duality group of string and brane theory with a non-zero $B$-field.
Furthermore, (3.3) for the irrational $\theta$ can be treated as the 
area-preserving diffeomorphism algebra of the NCT, SDiff(${T_{\theta}}^2$).
\newline
{\it The gauge theory action}
\newline
Now, we imply (3.1) in writing the action (2.18) in momentum space
in terms of $L_{\vec{k}}$ and Fourrier modes. Here we perform
calculations with the non-supersymmetric action, its generalization to
the supersymmetric case, obtained from (2.20), is like the usual gauge theories,
thus we only quote the results.

Plugging  (3.1) into the (2.17), we have
\be
F_{\mu\nu}= \sum_{\vec{k}} -ik_{[\mu}A_{\nu]}(k) L_{\vec{k}} +
2 i\sum_{\vec{m},\vec{n}} \sin (\pi\theta\vec{m}\times\vec{n})
A_{\mu}(m)A_{\nu}(n) L_{\vec{m}+\vec{n}}
\ee
and hence the action of the $U(1)$ NCYM, (2.18), in Fourrier modes is:
\be\bea{cc}
S={1\over g^2_{YM}} \sum_{\vec{k}} \bigl(-k_{[\mu}A_{\nu]}(k)\bigr)
\bigl(k_{[\mu}A_{\nu]}(-k)\bigr)+\\
2 \sum_{\vec{m},\vec{n},\vec{k}} k_{[\mu}A_{\nu]}(-k)\sin(\pi\theta\vec{m}
\times\vec{n})A_{\mu}(m)A_{\nu}(n) \delta(\vec{m}+\vec{n}+\vec{k})+\\
4 \sum_{\vec{m},\vec{n},\vec{k},\vec{l}}\sin(\pi\theta\vec{m}\times\vec{n})
\sin(\pi\theta\vec{k}\times\vec{l})A_{\mu}(m)A_{\nu}(n)
A_{\mu}(-k)A_{\nu}(-l) \delta(\vec{m}+\vec{n}+\vec{k}+\vec{l}).
\eea\ee
In writing the above action we have implied the reality condition of the
fields, $A_{\mu}(-k)=A_{\mu}^*(k)$.
The above action is of the form of the usual gauge theories in which group
indices are identified with the momentum, the structure constants
$f_{\vec{m},\vec{n},\vec{k}}$ are
\be
f_{\vec{m},\vec{n},\vec{k}}=2 \sin(\pi\theta\vec{m}\times\vec{n})
\delta(\vec{m}+\vec{n}+\vec{k})
\ee
As we see the momentum conservation condition is automatically taken into account
by the above structure constants. The important feature of this gauge theory
is  {\it gauge invariance}, (2.19), which enables us to handle the
problem of renormalizability without caring about the non-locality of the theory.

Noting the (3.7) and the action (3.6), we see that in this gauge theory our
momentum dependent coupling is $\sin(\pi\theta\vec{m}\times\vec{n})$, which is
smaller than one, hence we hope that the usual perturbative field theory
methods, despite the momentum dependence of the couplings, work here.

Another point to mention here is the broken Lorentz invariance of (2+1)
dimensional space. As it is seen from the action the couplings are a
function of the {\it spatial} momentum only, which explicitly breaks
the Lorentz invariance. The gauge fixing condition removes the subtleties
related to the lack of Lorentz invariance. We will show later that, this
Lorentz non-invariance  due to the special form of the  interactions will not
destroy our perturbative calculations and is consistent with the gauge
invariance at least, at one loop. Moreover we can find a "Ward identity"
for this gauge theory.

Before going to the details of loop calculations, let us study the structure
of the algebra defined by (3.3) and the related gauge theory, NCYM:

a) All the  $L_{\vec{m}}$  with parallel $\vec{m}$ are commuting.

b) The only generator which commutes with all the others is $L_{\vec{0}}$,
so rank of the group is one and the related Cartan sub-algebra is $U(1)$.

c) In the usual non-Abelian gauge theories, the gauge particles are charged
under the Cartan sub-algebra, in our case since the Cartan is $U(1)$, the
gauge particles are "{\it dipoles}" under the sub-algebra [19], and
the corresponding dipole moments are proportional to their {\it momentum}
but always {\it normal} to it and parallel dipoles are non-interacting.

d) Because the momentum is conserved in each vertex, here we have 
the {\it dipole moment conservation}.

e) From the string theory point of view [19], these dipoles are lowest modes 
of the strings attached to a D2-brane in a B-field background.

f) The high energy dipole-dipole scattering is suppressed by the  
Moyal bracket structure.
\newline
{\it Perturbative tools}
\newline
To do the calculations we need to read off the propagators and interaction
vertices from the action (3.6). For further details we refer the reader to [20].

\begin{figure}[h]
\begin{picture}(100,220)(70,90)
\epsfbox{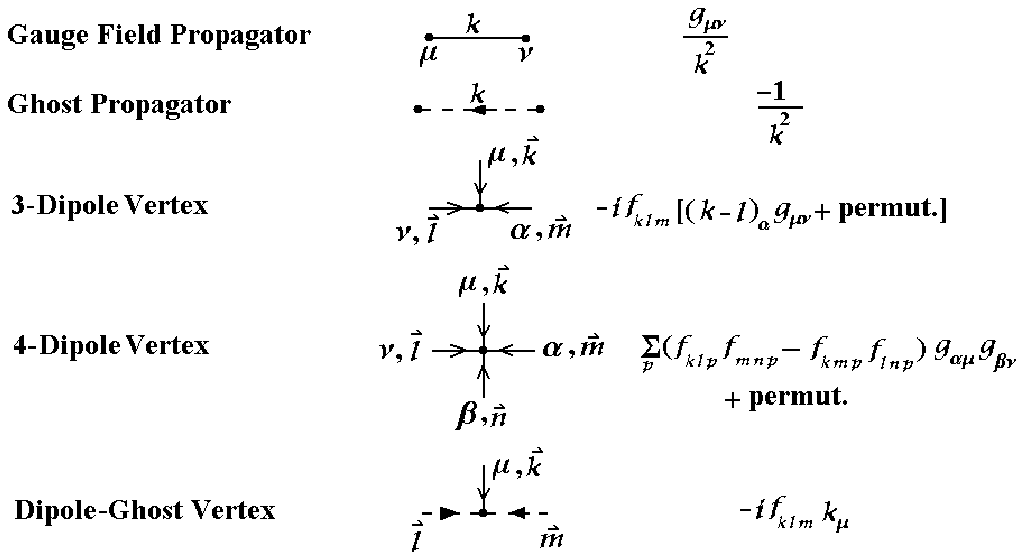}
\end{picture}
\end{figure}

\section{Loop Calculations and Renormalization}
\setcounter{equation}{0}

In this section, in order to discuss the renormalizability,
$\beta$-function behaviour of the theory and its UV and IR limits, 
we explicitly calculate the one loop diagrams of the NCYM.
For simplicity, we take the continuum limit and instead of
summation we use integrals. The theory is supposed to have a well-defined
continuum limit [13].
\newline
{\it Loop calculations}
\newline
1) The dipoles self energy part:
\newline
It gets contributions from four diagrams, dipole-loop, dipoles tadpole-loop,
ghost-loop and the counter-term [20].
\newline
1-1) Dipole-loop:
\be
\Pi_{\mu\nu}(k)={1\over 2}\int {d^3 q \over(2\pi)^3}{1\over q^2(q+k)^2}A_{\mu\nu}
\bigl(2\sin(\pi\theta \vec{k}\times\vec{q})\bigr)^2,
\ee
with
\be
A_{\mu\nu}=(q^2+(q+k)^2+4k^2)g_{\mu\nu}+6q_{\mu}q_{\nu}+3(q_{\mu}k_{\nu}+
q_{\nu}k_{\mu})-3k_{\mu}k_{\nu}.
\ee
It is worth noting that, in the case of NCYM theories, unlike the usual
loop calculations, since the theory is not Lorentz invariant, we can not
use the Wick rotation method.
To perform the loop integrals, since there is no $q_0$ dependence in the vertex 
functions, first we integrate over the $q_0$ component by taking the residue 
of $q_0$ poles. The remaining integrals are of the form:
\be
\Pi_{\mu\nu}(k)={4\over 2}\int {qdqd\phi \over(2\pi)^3} 
\bigl[{A_{\mu\nu}|_{q_0=|q|}\over (k^2+2kq\cos \phi)}-
{A_{\mu\nu}|_{q_0=-|q|}\over (k^2-2kq\cos \phi)}\bigr]
\sin(\pi\theta kq\sin\phi)^2,
\ee
Before performing  the integration over $\phi$, we note that 
$$\sin^2(\pi kq\theta\sin\phi)=
{1\over 2}-{1\over 2}\cos(2\pi\theta kq \sin\phi).
$$
The cosine part can be expanded in terms of Bessel functions [21]. It is a
straightforward, but messy calculation to show that all the terms 
containing Bessel functions are finite, and the only divergent part 
comes from the ${1\over 2}$ term of the sine squared. 
So, to find the divergent part of the integral it is
enough to substitute the sine squared in (4.1) for ${1\over 2}$. In other
words, the divergent part of (4.1) has exactly the form of the usual $SU(N)$
YM theory with $C_2(G)=4\times{1\over 2}$. 
Furthermore, the divergent part of 
(4.1) is {\it Lorentz invariant} and can be calculated by Wick rotation.
\newline
1-2) Dipole tadpole- and Ghost- loops:
\newline
Using the method explained earlier, i.e. integrating on $q_0$ component and
expanding the sine squared in terms of Bessel functions, one can show that
the divergent part of these diagrams, like the dipole-loop case gets 
contributions only from putting the sine squared equal to ${1\over 2}$, which
again restores the {\it Lorentz invariance} of the theory.

So altogether the dipole self energy part, is renormalized like a usual 
local gauge theory, with $C_2(G)=2$. The other important result is that 
although the interactions introduced by NCYM theory are not Lorentz invariant, 
the Lorentz invariance {\it remains} in the propagators at one loop.
\newline
2) Ghost self energy part:
\newline
There are only two diagrams contributing here, the ghost-dipole loop and the
counter-term. The integrals appearing in ghost-dipole loop, is similar to 
the ghost-loop contributions discussed above, hence our argument holds 
true and the divergent part is obtained from the usual $SU(N)$ gauge theory
with $C_2(G)=2$.

The other important result we obtain here is a generalized "Ward identity",
$$
k^{\mu}\Pi_{\mu\nu}=0.
$$
\newline
3) dipole-ghost vertex:
\newline
There are three diagrams contributing here, the diagram with two-ghosts in the
loop, the diagram with one ghost in the loop and the counter-term. 
Performing calculations by the method explained above, after a long but 
straight-forward  calculus, we again find that the {\it divergent part} of the 
loops is described by usual gauge theory result with $C_2(G)=2$.

For the three and four dipole vertices the same results are obtained.
We see from explicit loop calculations that the divergent part of the
NCYM theory at one loop, is governed by the divergent part of the loops of
a usual gauge theory. Hence the divergent parts behave like a local field
theory. However there are finite contributions at one loop level which 
are {\it non-local}. These terms also break the "Lorentz invariance". Appearance
of such terms suggests the possibility of non-local divergent terms
at higher loops [12]. But, according to our calculations [22], due to the
Bessel functions, the contribution of the non-local terms at {\it two loop}
are again suppressed.

We should remind here that, in all of the calculations we have assumed that 
the structure constant, $2 \sin(\pi\theta\vec{m}\times\vec{n})$, is non-zero.
So, although our result is $\theta$ independent, the above arguments
are  not valid for $\theta=0$ case 
\footnote{ In the rational $\theta$ case always one can find a basis on the
torus under which the structure constant is zero.}.

\section{ UV and IR Behaviour of Noncommutative Gauge Theories}
\setcounter{equation}{0}
By explicit loop calculations, it was shown in previous section that the 
NCYM theory is one loop renormalizable. Here we want to discuss the UV and IR
limits in more detail.
Since we consider the theory on a torus, however in the large volume limit,
$p$, the momentum, is discrete and we need not to address the 
$p\rightarrow 0$ behaviour. 

According to our results, although the theory is non-local and hence the 
usual renormalization group arguments are not valid, the one loop 
$\beta$-function of NCYM is negative, it is asymptotically free. 
Moreover in the large volume limit, the UV disentangles from the IR [13]. 

In the UV limit, besides the formal arguments of [13] our explicit 
calculations show that, we deal with a {\it non-local} theory. In UV, although
the theory is not scale invariant, it admits a fixed point.

The supersymmetric case, NCSYM, is realized by the D2-brane in a B-field 
background [2,11] which in the $\alpha'\rightarrow 0$, the low energy
dynamics, like the usual D-brane arguments, decouples from the bulk.
So the collective coordinates of the D2-brane, which are living in a commuting 
space, form a single vector multiplet. The 7 scalars correspond to 7 transverse
direction of the brane. The dual of the vector multiplet, which is a scalar
corresponds to the position of the brane in the eleventh dimension [23].
Hence the moduli space of the $U(1)$ NCSYM theory is $R^7\times S^1$. The
coupling of this (2+1) dimensional  gauge theory is given by the circumference 
of the $S^1$ factor. Besides the transverse coordinates, since we are 
interested 
in the gauge theories on NCT, there are moduli coming from the compact space 
and noncommutative structure of the tours. In the string theory limit, these
moduli are those the T-duality group acting on 
\be
{\cal M}_c={SL(2)_N \over SO(2)}\times {SL(2)_c \over SO(2)}.
\ee
The  $SL(2)_N$ acts on the deformation parameter of the torus and the other 
on the corresponding radii. Considering the contribution from the flat space
we have
\be
{\cal M}={\cal M}_c\times R^7\times S^1
\ee
At strong coupling the $S^1$ factor combines with $SL(2)_c$ and hence [8]
\be
{\cal M}={SL(2)_N \over SO(2)}\times {SL(3)_c \over SO(3)}\times R^7.
\ee

\section{Concluding remarks}
In this paper, we have studied the SYM on the NCT more extensively. The NCSYM 
is described by the usual gauge theory action, with the commutators substituted 
for the {\it Moyal bracket}. 
As discussed in [2,11], the NCSYM describes the low energy dynamics of D2-brane
wrapped on $T^2$ in a B-field background, and in $\alpha'\rightarrow 0$ it 
decouples from the bulk. Studying the NCT, we argued that by means of 
{\it Morita equivalence} we can map the $U(N)$ NCSYM with a magnetic flux $M$
to a $U(1)$ NCSYM theory, hence we only considered the $U(1)$ case in our later
arguments.

Using Fourrier mode expansion, we wrote the Moyal bracket algebra in terms of 
Fourrier basis, (3.3). The key idea, is the existence of commutative derivatives,
could be defined on the NCT. Hence one expects by going to the momentum
space, the
NCYM  translate to a commutative gauge theory. We argued that the gauge group
is the generalization of the area-preserving diffeomorphisms of the torus to the
NCT, $SDiff(T^2_{\theta})$. As discussed, this gauge group do not change
the 
Kahler structure of the NCT. Writing the NCYM action in the Fourrier modes, we 
showed that it is like a usual gauge theory, with the structure constants  
given by (3.7). The group indices are identified with the momentum. 

We showed that the Cartan sub-algebra in our case is $U(1)$, and hence
justified the string theoretic arguments of [19]. The other gauge particles
are "dipoles" under this $U(1)$ part.
Having the action, we worked out the loop calculations, and found out that

{\it i)} Although the NCYM is not a local gauge theory in the usual sense 
and the interactions introduced in it are not Lorentz invariant, the
divergent parts of the propagator loops are Lorentz invariant and the  
divergent part of the interaction vertices have the same structure
of the classical ones. This enables us to make the statements we already know 
about the usual gauge theories, for the NCYM too. The divergent parts
of the loops are given by the usual gauge theory results with the
quadratic Casimir equal to 2.
Moreover we can have generalized "Ward identities" here.

{\it ii)} The non-locality in the theory, is not removed in the loops, but
at one loop, since the theory is renormalizable, it has the same non-local 
structure. In other words the structure 
constants of the theory are {\it not renormalized}, like the usual gauge 
theories.
It seems that, the contribution of the non-local terms to the dipoles self energy 
are finite and also suppressed at high energies so that, they will not
give further contributions at higher loops [22].

{\it iii)} Since the loop corrections are similar to the usual $SU(N)$ 
gauge theories, the NCYM theories are {\it asymptotically} flat and
in spite of the non-locality, the theory has got a UV fixed point.

{\it iv)} NCYM are among the few gauge theories which admit the 
maximally super extension. As argued here, these theories are {\it exact}. 
The moduli space structure of these theories, compared with the usual
(2+1) ${\cal N}=8$ SYM, have an extra $Sl(2)$ factor acting on the 
noncommutative structure of the torus. Since this extra factor changes
the deformation parameter $\theta$, the structure constants of the theory
are also changed, but its behaviour is not altered. 

The possible singular structure of the moduli space, is an open question
we will postpone it to a future work.

When I was typing the manuscript, [24] appeared on the net in which
the same problem has been considered.

{\bf Acknowledgements}

I would like to thank K. Kaviani, for his collaboration at the early stages of
the work and reading the manuscript. I would like to thank H. Arfaei, F.
Ardalan and A. Fatolahi for many  fruitful discussions and comments.

\end{document}